\documentclass[pra,aps,twocolumn,superscriptaddress,showpacs,reprint]{revtex4}
\usepackage{amsfonts}
\usepackage{amsmath}
\usepackage{amssymb}
\usepackage{graphicx}
\usepackage{epstopdf}

\providecommand{\U}[1]{\protect\rule{.1in}{.1in}}

\input{tcilatex}

\begin{document}

\preprint{HEP/123-qed}
\title[ ]{Squeezing effect induced by minimal length uncertainty}
\author{Yue-Yue Chen}
\affiliation{State Key Laboratory of High Field Laser Physics, Shanghai Institution of
Optics and Fine Mechanics, Chinese Academy of Sciences, Shanghai 201800,
China}
\author{Xun-Li Feng}
\email{xlfeng@siom.ac.cn}
\affiliation{State Key Laboratory of High Field Laser Physics, Shanghai Institution of
Optics and Fine Mechanics, Chinese Academy of Sciences, Shanghai 201800,
China}
\affiliation{Centre for Quantum Technologies, National University of Singapore, 2 Science
Drive 3, Singapore 117542}
\author{C. H. Oh}
\affiliation{Centre for Quantum Technologies, National University of Singapore, 2 Science
Drive 3, Singapore 117542}
\author{Zhi-Zhan Xu}
\email{zzxu@mail.shcnc.ac.cn}
\affiliation{State Key Laboratory of High Field Laser Physics, Shanghai Institution of
Optics and Fine Mechanics, Chinese Academy of Sciences, Shanghai 201800,
China}

\begin{abstract}
In this work, the dynamics of the deformed one-dimensional harmonic
oscillator with minimal length uncertainty is examined and the analytical
solutions for time evolution of position and momentum operators are
presented in which the rough approximation that neglects the higher order
terms in BakerHausdor lemma is avoided. Based on these analytical solutions
the uncertainties for position and momentum operators are calculated in a
coherent state, and an unexpected squeezing effect in both coordinate and
momentum directions is found in comparison with ordinary harmonic
oscillator. Obviously such a squeezing effect is induced by the minimal
length uncertainty (gravitational effects). Our results are applied to the
electrons trapped in strong magnetic fields to examine the degree of the
existing squeezing effect in a real system, which shows the squeezing degree
induced by minimal length uncertainty is very small.
\end{abstract}

\volumeyear{year}
\volumenumber{number}
\issuenumber{number}
\eid{identifier}
\date[Date text]{date}
\received[Received text]{date}
\revised[Revised text]{date}
\accepted[Accepted text]{date}
\published[Published text]{date}
\startpage{1}
\endpage{102}
\maketitle

\renewcommand\thesection{\arabic{section}}

\section{I\MakeLowercase{ntroduction}}

The general relativity and quantum mechanics are expected to be unified at
Plank scale. The theories that seek to formulate a quantum theory of
gravitation are termed as quantum gravity. Almost all the promising
approaches to quantum gravity, such as string theory \cite{String}, loop
quantum gravity \cite{loop}, and black hole physics \cite{black hole}
suggest the existence of a minimum measurable length. However, such minimal
length scale is obviously against the Heisenberg principle $\Delta x\Delta
p\geq \hslash /2$, in a way that, arbitrarily precise measurement of
position $x$ is no longer possible even at the cost of our knowledge about $%
p $. To take into account the effect of minimal length on quantum mechanics,
many proposals for quantum gravity suggested a generalized Heisenberg
uncertainty principle (GUP) that deforms the canonical one to \cite{Saurya}%
\begin{equation}
\Delta \hat{x}\Delta \hat{p}\geq \hbar /2(1+\beta \Delta \hat{p}^{2}),
\end{equation}%
where $\beta =\beta _{0}/(M_{Pl}c)^{2}=l_{Pl}^{2}/2\hbar ^{2},\beta _{0}$ is
a parameter that quantifies the correction strength, $l_{Pl}$ is \ the Plank
length. GUP implies the very existence of minimum measurable length
predicted by quantum gravity $\Delta x_{\min }=\hslash \sqrt{\beta }$ below
which $\Delta x$ cannot be reduced. Specifically, any quantum state with a
position uncertainty $\Delta x$ inside the region $\Delta x<\Delta x_{\min }$
was proved to have infinite energy, and the expectation value of kinetic
energy $\left\langle E_{p}\right\rangle $ in such states is actually
divergent regardless of the representation adopted \cite{Kempf}. The
emergence of a minimal length scale and GUP will modify all the quantum
system with a well-defined Hamiltonian, including low-energy quantum
mechanics phenomena. Depending on the value of the dimensionless parameter $%
\beta _{0},$ the unobservable corrections can be interpreted in two
different ways. Specifically, one can adopt the regular assumption that $%
\beta _{0}$ is of the order of unity, in which case the corrections are
trivial unless energies (lengths) are near Planck energy (length), or
alternatively, set an intermediate length scale between electron weak length
scale and Plank scale with a much larger $\beta _{0}$ ($\beta _{0}\gg 1$) 
\cite{Saurya,Ali}. Recently, GUP has been extensively studied and numerous
proposals have been made to recheck the various quantum systems in the
context of GUP \cite{KT,MV,B,K,PP}. It also offers an alternative way to
explore quantum gravity effects in terms of measurement of deviation from
standard quantum mechanics due to GUP \cite{FM,Igor}.

Harmonic analysis has appeared in a vast range of approaches and techniques
in quantum mechanics and quantum optics. Because of its importance as a
basic study model, many efforts have been devoted to this subject and the
relevant theories are developed to maturity under canonical quantum
mechanics frame \cite{Harmonic}. Recently, the studies of noncommutative
spacetime structures have injected new vitality in this field. Harmonic
analysis with minimum length scale can be used as an elementary input for
many techniques to address Plank-scale physics using quantum mechanics and
quantum optics \cite{AKF,AK,CQ,Al,Igor,BM,PPP,SD,CLC}. Considerable physical
problems can be regarded as a deformed harmonic oscillator with GUP, such as
the oscillations of a carbon monoxide molecule \cite{BM}, the Landau
problem, ultra cold neutrons bouncing above a mirror \cite{K}, and singular
Calogero potential in one dimension \cite{Saurya,Polychronakos}. The problem
of harmonic oscillator with the minimal length uncertainty has been
considered previously by Kempf et al. \cite{Kempf}. They presented the
analytical solution of harmonic oscillator by solving Schr\"{o}dinger
equation, in terms of the energy eigenvalues and eigenstates. Refs. \cite%
{Zachary,Lay} further generalized the analytical result from the
one-dimensional to D-dimensional harmonic oscillator. Approaches to
construct generalized coherent states for harmonic oscillator in the
deformed quantum mechanics have been proposed \cite{P,Ching}. Ghosh et al.
obtained the $\hat{x}$ and $\hat{p}$ uncertainty for the generalized
coherent states \cite{Antoine} for a generalized harmonic oscillator, and
verified the GUP \cite{Subir}. However, the results were restricted to a
well-designed coherent state with four requirements including temporal
stability, which moderated the importance of studies on time evolution. The
evolution of position and momentum operators in the context of GUP has been
discussed in \cite{PPP} with classical description. The quantum description
has been attempted in \cite{Kourosh}. However, the solutions given were
based on the approximation that neglects all the higher order terms. The
feasibility of the approximation will be broken down at longer time or
larger frequency.

In this paper, we first give the analytical expressions for time dependent
canonical operators of the deformed harmonic oscillator in section II. In
section III, based on the results obtained, we present the analytical
solution for time evolution of deformed operators $\hat{x}(t)$, $\hat{p}(t)$
and their uncertainties in a coherent state. Then, we study the temporal
behavior of the those uncertainty with the normal quantum mechanics case.
Surprisingly, an squeezing effect appears in both position and momentum
directions. To further understand this squeezing effect, we use the
parameters of electrons trapped in strong magnetic fields to evaluate the
significance and magnitude of this effect. In section IV, we end with a
conclusion.

\section{T\MakeLowercase{ime evolution of canonical operators for deformed
harmonic
oscillations by} GUP}

The GUP is equivalent to the following modified commutator \cite{Ivan}:%
\begin{equation}
\lbrack \hat{x},\hat{p}]=i\hbar (1+\beta \hat{p}^{2}).
\end{equation}%
In this paper, we consider the one-dimensional Darboux map \cite{K},%
\begin{equation}
\hat{x}=x,\hat{p}=p(1+\frac{\beta }{3}p^{2}),
\end{equation}%
where operators with a hat denote the present deformed operators otherwise
denote canonical operators. Apparently Eq. (2) is satisfied to $O(\beta )$
with the map (3), thus we neglect terms higher than order $\beta $ to keep
only $O(\beta )$ correction throughout this paper. The Hamiltonian of
harmonic oscillator thus becomes%
\begin{equation}
H=\frac{\hat{p}^{2}}{2m}+\frac{m\omega ^{2}\hat{x}^{2}}{2}=\frac{p^{2}}{2m}+%
\frac{m\omega ^{2}x^{2}}{2}+\frac{\beta p^{4}}{3m}.
\end{equation}%
For our convenience, we introduce canonical annihilation and creation
operators%
\begin{equation}
a=\sqrt{\frac{m\omega }{2\hbar }}(x+\frac{ip}{m\omega }),a^{\dagger }=\sqrt{%
\frac{m\omega }{2\hbar }}(x-\frac{ip}{m\omega }).
\end{equation}%
The Hamiltonian is then rewritten as%
\begin{equation}
H=\hbar \omega (a^{\dagger }a+\frac{1}{2})+\frac{1}{12}\hbar ^{2}\omega
^{2}m\beta (a-a^{\dagger })^{4}
\end{equation}%
In Heisenberg picture of quantum mechanics, motion equation for canonical
annihilation operator $a$ is $\frac{da}{dt}=\frac{i}{\hbar }[H,a].$ The time
evolution of $a$ is obtained by using Baker-Hausdorff lemma (the lengthy
calculation is shown explicitly in the Appendix A). 
\begin{eqnarray}
a(t) &=&e^{\frac{i}{\hbar }Ht}ae^{-\frac{i}{\hbar }Ht}  \notag \\
&=&ae^{-i\omega t}+\hbar \omega m\beta \lbrack -i\omega te^{-i\omega
t}(a+a^{\dagger }a^{2})+  \notag \\
&&i\sin \omega t(a^{\dagger }+a^{\dagger 2}a)+\frac{1}{6}(e^{-i\omega
t}-e^{-3i\omega t})a^{3}  \notag \\
&&+\frac{1}{12}(e^{-i\omega t}-e^{3i\omega t})a^{\dag 3}],
\end{eqnarray}%
and creation operator $a^{\dagger }(t)=(a(t))^{\dagger }$. Now let us
introduce the following two canonical quadrature operators 
\begin{equation}
X_{1}=\frac{1}{2}(a(t)+a^{\dagger }(t)),\text{ }X_{2}=\frac{1}{2i}%
(a(t)-a^{\dagger }(t)).
\end{equation}%
Their variances are defined as 
\begin{eqnarray}
(\Delta X_{1})^{2} &=&\left\langle X_{1}^{2}\right\rangle -(\left\langle
X_{1}\right\rangle )^{2},  \notag \\
(\Delta X_{2})^{2} &=&\left\langle X_{2}^{2}\right\rangle -(\left\langle
X_{2}\right\rangle )^{2}.
\end{eqnarray}%
\newline
We proceed to compute the product of quadrature operators uncertainty $%
(\Delta X_{1})^{2}(\Delta X_{2})^{2}$ in coherent state $\left\vert \alpha
\right\rangle ,$ which is the eigenstate of canonical annihilation operator $%
a$. The computation results are given below (see Appendix B for more
details),%
\begin{equation}
\begin{split}
(\Delta X_{1})^{2}=& \frac{1}{4}+\frac{1}{16}e^{-2i\omega t}\hbar \omega
m\beta \lbrack (-2+3\alpha ^{2}-4i\omega t\alpha ^{2} \\
& +\alpha ^{\ast 2}-8\alpha \text{Re}[\alpha ])+4e^{2i\omega
t}(1+2\left\vert \alpha \right\vert ^{2}) \\
& +e^{4i\omega t}(-2+\alpha ^{2}-4\left\vert \alpha \right\vert
^{2}+4i\omega t\alpha ^{\ast 2}-\alpha ^{\ast 2})],
\end{split}%
\end{equation}%
\begin{equation}
\begin{split}
(\Delta X_{2})^{2}=& \frac{1}{4}+\frac{1}{16}e^{-2i\omega t}\hbar \omega
m\beta \lbrack (2-3\alpha ^{2}+4i\omega t\alpha ^{2} \\
& -\alpha ^{\ast 2}+8\alpha \text{Re}[\alpha ])-4e^{2i\omega
t}(1+2\left\vert \alpha \right\vert ^{2}) \\
& +e^{4i\omega t}(2-5\alpha ^{2}+4i\omega t\alpha ^{\ast 2}-\alpha ^{\ast
2})].
\end{split}%
\end{equation}%
Note that $X_{1}$ and $X_{2}$ are canonical operators with commutation
relation and are supposed to fit the feature of ordinary quantum mechanics
regardless of unitary transformation. Thus, the minimum uncertainty
requirement of coherent state $(\Delta X_{1})^{2}(\Delta X_{2})^{2}=\frac{1}{%
16}$ can be considered as a criterion to justify our computation results to
some extent. The product of Eq. (10) and Eq. (11) can exactly meet the
requirement.

\section{s\MakeLowercase{queezing effect induced by} GUP}

With the solution for canonical operators given in Appendix A, it is easy to
obtain the time evolution of position and momentum in the framework of GUP
with the mapping Eq. (3), 
\begin{widetext}
\begin{eqnarray}
\hat{x}(t) &=&x(t)=\sqrt{\frac{2\hbar }{m\omega }}X_{1} \notag\\
&=&\sqrt{\frac{\hbar }{2m\omega }}(ae^{-i\omega t}+a^{\dagger }e^{i\omega
t})+\frac{\beta e^{-3i\omega t}}{12}\sqrt{\frac{\hbar ^{3}m\omega }{2}}%
[-6e^{2i\omega t}(-1+e^{2i\omega t}+2it\omega )a+  \notag \notag\\
&&12ie^{3i\omega t}a^{\dagger }(e^{i\omega t}t\omega +\sin \omega
t)+(2e^{2i\omega t}-3+e^{4i\omega t})a^{3}-(12ie^{2i\omega t}t\omega
+12ie^{3i\omega t}\sin \omega t)a^{\dagger }a^{2} \notag\\
&&+(12ie^{4i\omega t}t\omega +12ie^{3i\omega t}\sin \omega t)a^{\dagger
2}a)+(e^{2i\omega t}+2e^{4i\omega t}-3e^{6i\omega t})a^{\dagger 3}],
\end{eqnarray}

\begin{eqnarray}
\hat{p}(t) &=&p(t)(1+\frac{\beta p^{2}(t)}{3})=\sqrt{2\hbar m\omega }(X_{2}+%
\frac{2}{3}\hbar m\omega \beta X_{2}^{3}) \notag\\
&=&i\sqrt{\frac{\hbar m\omega }{2}}(a^{\dagger }e^{i\omega t}-ae^{-i\omega
t})-\frac{1}{12\sqrt{2}}e^{-3i\omega t}(\hbar m\omega )^{3/2}\beta \lbrack
6e^{2i\omega t}(ie^{2i\omega t}+2t\omega )a  \notag \notag\\
&&+6e^{2i\omega t}(-i+2e^{2i\omega t}t\omega )a^{\dagger }+(6ie^{4i\omega
t}+12e^{2i\omega t}t\omega )a^{\dagger }a^{2}+(-6ie^{2i\omega
t}+12e^{4i\omega t}t\omega )a^{\dagger 2}a \notag\\
&&+i(-3+2e^{2i\omega t}-e^{4i\omega t})a^{3}+i(e^{2i\omega t}-2e^{4i\omega
t}+3e^{6i\omega t})a^{\dagger 3}].
\end{eqnarray}%
\end{widetext}Obviously, Eq. (12) and Eq. (13) return to normal case with
vanishing $\beta .$ The dynamics is neither periodic nor harmonic due to the
influence of GUP. Note that, the solutions obtained for time evolution of $%
\hat{x}(t)$ and $\hat{p}(t)$ are different from that given in \cite{Kourosh}
even expressed in terms of $\hat{x}(0)$ and $\hat{p}(0)$ using Eq. (5). In
the framework of fist order correction, the results presented in this paper
are much more precise. We avoid the controversial approximation adopted in 
\cite{Kourosh} that neglects the terms of order $(\omega t)^{5}$ and higher.
However, our numerical results show that such an approximation is incorrect
for a relatively large $\omega t$ because in a term with higher order
exponential of $\omega t$ the increase of the exponential part may overwhelm
the decrease of its coefficient, thus the whole term may play a more
important role than the term with the lower order exponential of $\omega t$,
thus cannot be neglected any more. Next, we proceed to calculate the $\hat{x}
$-variance and $\hat{p}$-variance in coherent state.%
\begin{equation}
(\Delta \hat{x})^{2}=(\Delta x)^{2}=\frac{2\hbar }{m\omega }(\Delta
X_{1})^{2},\text{ }
\end{equation}%
\begin{equation}
(\Delta \hat{p})^{2}=2\hbar \omega m(\Delta X_{2})^{2}+\frac{8}{3}\hbar
^{2}\omega ^{2}m^{2}\beta (\left\langle X_{2}^{4}\right\rangle -\left\langle
X_{2}\right\rangle \left\langle X_{2}^{3}\right\rangle ).
\end{equation}%
The derivation for $(\Delta \hat{x})^{2}$ is straightforward, while to get $%
(\Delta \hat{p})^{2}$ we have to calculate $\left\langle
X_{2}^{4}\right\rangle $ and $\left\langle X_{2}^{3}\right\rangle .$ The
calculation results are complicated and lengthy and here we only directly
give the results

\begin{equation}
\begin{split}
(\Delta \hat{x})^{2}=& \frac{\hbar }{2m\omega }+\frac{1}{8}e^{-2i\omega
t}\hbar ^{2}\beta \lbrack (-2+3\alpha ^{2}-4i\omega t\alpha ^{2} \\
& +\alpha ^{\ast 2}-8\alpha \text{Re}[\alpha ])+4e^{2i\omega
t}(1+2\left\vert \alpha \right\vert ^{2}) \\
& +e^{4i\omega t}(-2+\alpha ^{2}-4\left\vert \alpha \right\vert
^{2}+4i\omega t\alpha ^{\ast 2}-\alpha ^{\ast 2})]],
\end{split}%
\end{equation}%
\begin{eqnarray}
(\Delta \hat{p})^{2} &=&\frac{\hbar m\omega }{2}+\frac{1}{8}e^{-2i\omega
t}\hbar ^{2}\omega ^{2}m^{2}\beta \lbrack 2-7\alpha ^{2}+4i\omega t\alpha
^{2}  \notag \\
&&-\alpha ^{\ast 2}+8\alpha \text{Re}[\alpha ]+e^{4i\omega t}(2-5\alpha
^{2}-3\alpha ^{\ast 2}  \notag \\
&&-4i\omega t\alpha ^{\ast 2}+8\alpha \text{Re}[\alpha ])].
\end{eqnarray}%
Then the product of the coordinate and momentum variances follows 
\begin{equation}
\begin{split}
& (\Delta \hat{x})^{2}(\Delta \hat{p})^{2} \\
& =\frac{\hbar ^{2}}{4}+\frac{1}{4}\hbar ^{3}m\omega \beta (1-e^{-2i\omega
t}\alpha ^{2}+2\left\vert \alpha \right\vert ^{2}-e^{2i\omega t}\alpha
^{\ast 2}).
\end{split}%
\end{equation}%
Note that, the generalized Heisenberg uncertainty principle requires $%
(\Delta \hat{x})^{2}(\Delta \hat{p})^{2}\geq \frac{1}{4}\left\vert
\left\langle \left[ \hat{x},\hat{p}\right] \right\rangle \right\vert ^{2}$.
While 
\begin{equation}
\frac{1}{4}\left\vert \left\langle \left[ \hat{x},\hat{p}\right]
\right\rangle \right\vert ^{2}=\frac{\hbar ^{2}}{4}(1+2\beta \left\langle 
\hat{p}^{2}\right\rangle ),
\end{equation}%
substituting the expression of $\left\langle \hat{p}^{2}\right\rangle $ into
Eq. (19) yields exactly the Eq. (18). For the specific expression of $%
\left\langle \hat{p}^{2}\right\rangle ,$ we refer the readers to Appendix B.
So we conclude $(\Delta \hat{x})^{2}(\Delta \hat{p})^{2}=\frac{\hbar ^{2}}{4}%
(1+2\beta \left\langle \hat{p}^{2}\right\rangle ),$ that is, the deformed
harmonic oscillator with GUP in coherent states still satisfies the minimum
uncertainty relation to the first order correction.

To further show how the minimal length affects the dynamics of harmonic
oscillator, we compare the results in the framework of GUP with a simple
harmonic oscillator in the context of ordinary quantum mechanics with the
position and momentum operators denoted by $x_{0}$ and $p_{0}$ respectively.
The variances of canonical operators $x_{0}$ and $p_{0}$ are $(\Delta
x_{0})^{2}=\frac{\hbar }{2m\omega }$, $(\Delta p_{0})^{2}=\frac{\hbar
m\omega }{2}$, and their product is $(\Delta x_{0})^{2}(\Delta p_{0})^{2}=%
\frac{\hbar ^{2}}{4}$. This is actually the Eq.(16)-(18) when $\beta =0.$
Assuming eigenvalue $\alpha =\gamma e^{i\theta },$ where $\gamma $ is
positive and $\theta \in \lbrack 0,2\pi ],$ we calculate the difference
between the variances in GUP context and the corresponding canonical
variances.%
\begin{equation}
\begin{split}
& (\Delta \hat{x})^{2}-(\Delta x_{0})^{2} \\
& =\frac{\hbar ^{2}\beta }{2}[2\sin ^{2}\omega t+\gamma ^{2}(4\sin
^{2}\omega t-2\omega t\cos 2\theta \sin 2\omega t \\
& +(2\omega t\cos 2\omega t-\sin 2\omega t)\sin 2\theta )],
\end{split}%
\end{equation}%
\begin{equation}
\begin{split}
& (\Delta \hat{p})^{2}-(\Delta p_{0})^{2} \\
& =\frac{1}{4}\hbar ^{2}m^{2}\omega ^{2}\beta \lbrack 2\cos 2\omega t+\gamma
^{2}(4\cos 2\omega t-3\cos (2\omega t-2\theta ) \\
& -\cos (2\omega t+2\theta )+4\omega t\sin (2\omega t-2\theta ))],
\end{split}%
\end{equation}%
\begin{equation}
(\Delta \hat{x})^{2}(\Delta \hat{p})^{2}-(\Delta x_{0})^{2}(\Delta
p_{0})^{2}=1+2\gamma ^{2}(1-\cos (2\omega t-2\theta )).
\end{equation}%
Clearly, Eq. (22) indicates the product $(\Delta \hat{x})^{2}(\Delta \hat{p}%
)^{2}$ is always larger than $(\Delta x_{0})^{2}(\Delta p_{0})^{2}$
irrespective of the value of parameters, which is demanded by GUP.
Nevertheless, Eq. (20) and Eq. (21) cannot guarantee that the deformed
variances are invariably larger than their ordinary correspondence.
Specifically, by fixing $\gamma $, which generally has an overall influence
on amplitude, we plot the results of Eq. (20) and (21) as a function of
parameters $\omega t$ and $\theta $ in Fig. 1 and Fig. 2, respectively,
where the coefficients containing $\beta ,\hbar $ and $m$ outside bracket in
the r.h.s. are incorporated into the vertical coordinate,\ considering they
are positive and do not affect the sign of the outcome. Surprisingly, Fig. 1
and Fig. 2 show that the deformed variances $(\Delta \hat{x})^{2}$ and $%
(\Delta \hat{p})^{2}$ can be smaller than $(\Delta x_{0})^{2}$ and $(\Delta
p_{0})^{2}$, respectively. That is to say, squeezing effect unexpectedly
emerges in both $x$ and $p$ direction at certain time during the evolution. 
\begin{figure*}[tbp]
\begin{center}
\includegraphics[width=0.75\textwidth]{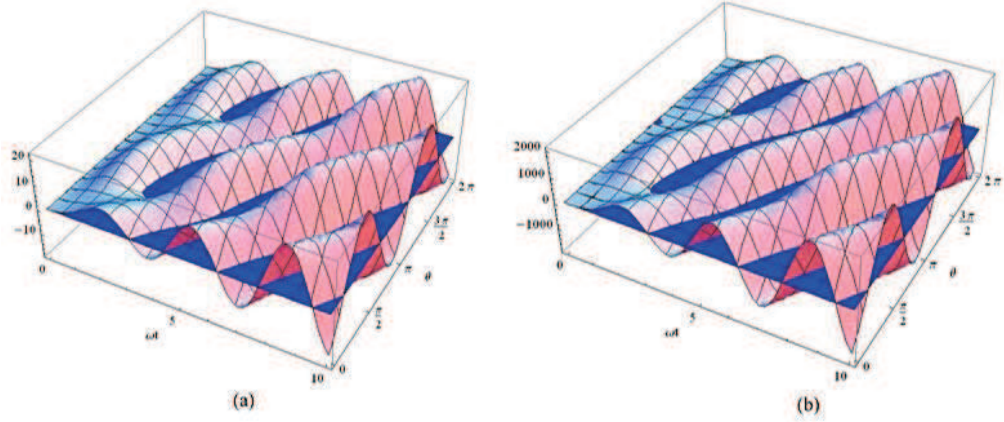}
\end{center}
\caption{(color online) The variation of $\frac{2}{\hbar ^{2}\protect\beta }%
[(\Delta \hat{x})^{2}-(\Delta x_{0})^{2}]$ with $\protect\omega t$ and $%
\protect\theta .$ The part that the difference below zero plane is where $%
(\Delta \hat{x})^{2}<(\Delta x_{0})^{2}$, which implicates the appearance of
squeezing. $\protect\gamma =1$ and $\protect\gamma =10$ in $(a)$ and $(b)$
respectively.}
\end{figure*}
\begin{figure*}[tbp]
\begin{center}
\includegraphics[width=0.75\textwidth]{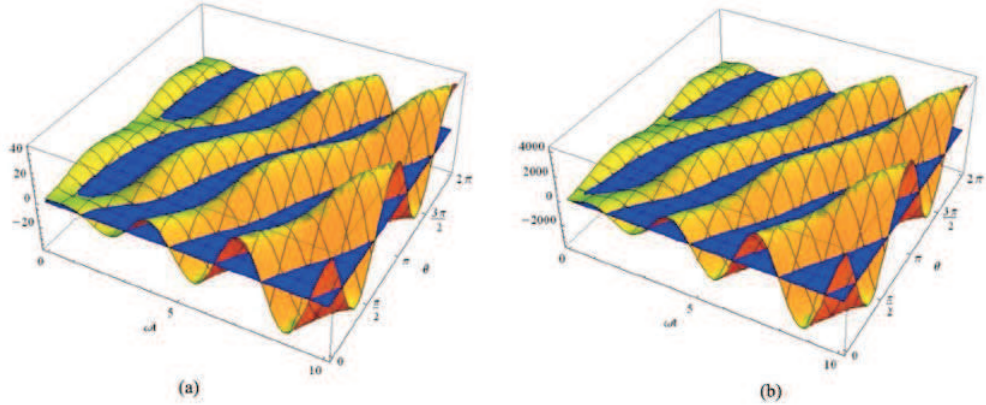}
\end{center}
\caption{(color online) The evolution of $\frac{4}{\hbar ^{2}m^{2}\protect%
\omega ^{2}\protect\beta }[(\Delta \hat{p})^{2}-(\Delta p_{0})^{2}]$ with $%
\protect\omega t$ and $\protect\theta .$ The part that the difference below
zero plane is where $(\Delta \hat{p})^{2}<(\Delta p_{0})^{2}$, which
implicates the appearance of squeezing. $\protect\gamma =1$ and $\protect%
\gamma =10$ in $(a)$ and $(b)$ respectively. }
\end{figure*}
To study the degree of such squeezing effect, now we apply our results to a
real system, an electron in a constant magnetic field. The cyclotron motion
of the electron is actually a harmonic oscillator. By measuring the energy
shift of Landau levels with a STM, an upper bound $\beta _{0}<10^{50}$ can
be defined \cite{Saurya}. The parameters are chosen as following: cyclotron
frequency $\omega _{c}\thickapprox 10^{3}$ GHz, $n=2,\alpha =e^{i\frac{\pi }{%
4}},$ and taking the largest allowed $\beta =\beta
_{0}/(M_{Pl}c)^{2}=2.43478\times 10^{48}.$ With those parameters, we are
able to give a more intuitive comparison between the standard and the
deformed harmonic oscillator based on uncertainty of position and momentum,
as depicted in Fig. 3. The picture shows that, for coherent state, both $%
(\Delta \hat{x})^{2}$ and $(\Delta \hat{p})^{2}$ oscillate, respectively,
around the straight lines of $(\Delta x_{0})^{2}=\frac{\hbar }{2m\omega }$
and $(\Delta p_{0})^{2}=\frac{\hbar m\omega }{2}$. Even though their product
always oscillates above the constant value of $(\Delta x_{0})^{2}(\Delta
p_{0})^{2}.$ That is, squeezing effect emerges in both $x$ and $p$
direction. But on the other hand, Fig. 3 (a) and (b) show that squeezing
degree for both $(\Delta \hat{x})^{2}$ and $(\Delta \hat{p})^{2}$ is very
small under the chosen parameters. Actually, a much smaller $\beta $ towards
Plank-scale modifications will render the squeezing effects negligible.    
\begin{figure*}[tbp]
\begin{center}
\includegraphics[width=0.95\textwidth]{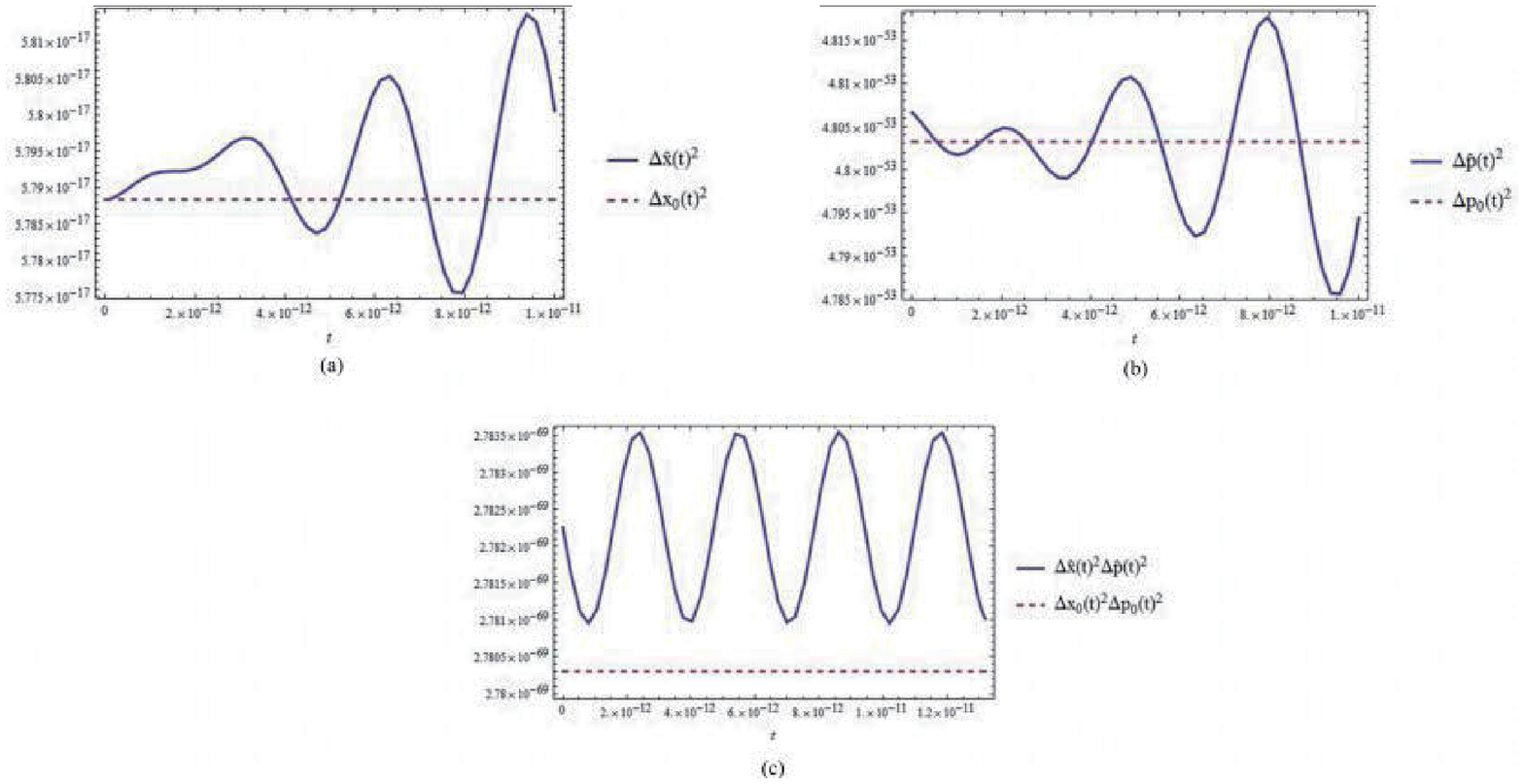}
\end{center}
\caption{(color online) The temporal behavior of variances of position $(a)$%
, momentum $(b)$ their product $(c)$ in a coherent state. The solid and
dashed lines respectively stand for deformed and canonical operators.}
\end{figure*}

\section{C\MakeLowercase{onclusion}}

In conclusion, we have presented the analytical expressions in Heisenberg
picture for time evolution of the operators of position $\hat{x}$ and
momentum $\hat{p}$ for the deformed harmonic oscillator with GUP, based on
which the uncertainties $(\Delta \hat{x})^{2}$ and $(\Delta \hat{p})^{2}$
for both position and momentum are calculated in a\ coherent state. We
surprisingly find $(\Delta \hat{x})^{2}$ and $(\Delta \hat{p})^{2}$ can be
smaller than that of a standard harmonic oscillator in a\ coherent state,
implying the minimal length uncertainty or gravitational effects can induce
squeezing effect in both position and momentum. As an example, we apply our
results to an electron trapped in strong magnetic fields by taking into
account influence of the minimal length uncertainty and find the existing
squeezing effect is actually very small. Our results may be useful for in
some techniques and approaches to explore and measure the potential quantum
gravitational phenomena with such as mechanical oscillator and trapped ions.

\section*{Acknowledgements}

Enlightened discussions with Prof. Yu Sixia are gratefully acknowledged.
This work was supported by NSFC (Grant Nos 11074079 and 11174081) and the
National Research Foundation and Ministry of Education, Singapore (Grant No.
WBS: R-710-000-008-271). \setcounter{equation}{0} \renewcommand{%
\theequation}{A\arabic{equation}} 
\begin{widetext}

\section*{Appendix A: Derivation for Eq.(7)}

The time evolution of $a$ can be obtained by using Baker-Hausdorff lemma
\begin{equation}
e^{\xi A}Be^{-\xi A}=B+\xi \lbrack A,B]+\frac{\xi ^{2}}{2!}[A,[A,B]]+......
\label{A1}
\end{equation}Let $\xi =\frac{i}{\hbar }t,$ $A=H,$ $B=a$, the second term of the
r.h.s. of Eq. (A1) is

\begin{equation}
\xi \lbrack A,B]=-it\omega a+it\hbar m\omega ^{2}\beta (-a+a^{\dagger }+%
\frac{1}{3}(a^{3}-a^{\dag 3})-a^{\dagger }a^{2}+a^{\dagger 2}a).
\end{equation}%
The third term is

\begin{equation}
\frac{\xi ^{2}}{2!}[A,[A,B]]=\frac{(it\omega )^{2}}{2!}a+\frac{(it)^{2}}{2!}%
\hbar m\omega ^{3}\beta (2a-\frac{4}{3}a^{3}-\frac{4}{3}a^{\dagger
3}+2a^{\dagger }a^{2}).
\end{equation}%
The fourth term is

\begin{equation}
\frac{\xi ^{3}}{3!}[A,[A,[A,B]]]=-\frac{(it\omega )^{3}}{3!}a+\frac{(it)^{3}%
}{3!}\hbar m\omega ^{4}\beta (-3a+a^{\dagger }+\frac{13}{3}a^{3}-\frac{7}{3}%
a^{\dagger 3}-3a^{\dagger }a^{2}+a^{\dagger 2}a).
\end{equation}%
The fifth term is

\begin{equation}
\frac{\xi ^{4}}{4!}[A,[A,[A,[A,B]]]]=\frac{(it\omega )^{4}}{4!}a+\frac{%
(it)^{4}}{4!}\hbar m\omega ^{5}\beta (4a-\frac{40}{3}a^{3}-\frac{20}{3}%
a^{\dagger 3}+4a^{\dagger }a^{2}).
\end{equation}%
The sixth term is

\begin{equation}
\frac{\xi ^{5}}{5!}[A,[A,[A,[A,[A,B]]]]]=-\frac{(it\omega )^{5}}{5!}a+\frac{%
(it)^{5}}{5!}\hbar m\omega ^{6}\beta (-5a+a^{\dagger }+\frac{121}{3}a^{3}-%
\frac{61}{3}a^{\dagger 3}-5a^{\dagger }a^{2}+a^{\dagger 2}a).
\end{equation}%
It can be seen that, coefficients of each term are actually terms of
expansion of a specific series and can be collected into a simple form. The
terms without contribution of $\beta $ is nothing but the usual results of
ordinary quantum mechanics,

\begin{equation}
1-it\omega +\frac{(it\omega )^{2}}{2!}-\frac{(it\omega )^{3}}{3!}+\frac{%
(it\omega )^{4}}{4!}-\frac{(it\omega )^{5}}{5!}+\ldots \ldots =\underset{n=0}%
{\overset{\infty }{\sum }}\frac{(-it\omega )^{n}}{n!}=e^{-i\omega t}.
\end{equation}%
The terms proportional to $\beta $ are induced by quantum gravitation.
Surprisingly, collection of coefficients for each operators can be
simplified either into exponential function or trigonometric function.
Specifically, collection of coefficients of $a+a^{\dagger }a^{2}$ has the
form

\begin{eqnarray}
&&hm\omega \beta (-it\omega +\frac{(it\omega )^{2}}{2!}\times 2-\frac{%
(it\omega )^{3}}{3!}\times 3+\frac{(it\omega )^{4}}{4!}\times 4-\frac{%
(it\omega )^{5}}{5!}\times 5+\ldots \ldots )  \notag \\
&=&\hbar m\omega \beta \underset{n=1}{\overset{\infty }{\sum }}\frac{%
(-it\omega )^{n}}{(n-1)!}=-it\hbar m\omega ^{2}\beta e^{-i\omega t}.
\end{eqnarray}%
For $a^{\dagger }+a^{\dagger 2}a,$ the result is

\begin{equation}
hm\omega \beta (it\omega +\frac{(it\omega )^{3}}{3!}+\frac{(it\omega )^{5}}{%
5!}+\ldots \ldots )=hm\omega \beta \underset{n=0}{\overset{\infty }{\sum }}%
\frac{(it\omega )^{2n+1}}{(2n+1)!}=i\hbar m\omega \beta \sin \omega t.
\end{equation}%
Collection of coefficients of $a^{3}$ turns out to be

\begin{eqnarray}
&&hm\omega \beta (it\omega \times \frac{1}{3}+\frac{(it\omega )^{2}}{2!}%
\times (-\frac{4}{3})+\frac{(it\omega )^{3}}{3!}\times \frac{13}{3}+\frac{%
(it\omega )^{4}}{4!}\times (-\frac{40}{3})+\frac{(it\omega )^{5}}{5!}\times
\frac{121}{3}+\ldots \ldots )  \notag \\
&=&-\hbar m\omega \beta \underset{n=1}{\overset{\infty }{\sum }}\frac{%
(-it\omega )^{n}}{n!}\times \frac{3^{n}-1}{3\times 2}=\hbar m\omega \beta
\frac{1}{6}(e^{-i\omega t}-e^{-3i\omega t}).
\end{eqnarray}%
Collection of coefficients of its complex conjugation term $a^{\dagger 3}$
is simplified to

\begin{eqnarray}
&&hm\omega \beta (it\omega \times (-\frac{1}{3})+\frac{(it\omega )^{2}}{2!}%
\times (-\frac{2}{3})+\frac{(it\omega )^{3}}{3!}\times (-\frac{7}{3})+\frac{%
(it\omega )^{4}}{4!}\times (-\frac{20}{3})+\frac{(it\omega )^{5}}{5!}\times
(-\frac{61}{3})+\ldots \ldots )  \notag \\
&=&-\hbar m\omega \beta \underset{n=1}{\overset{\infty }{\sum }}\frac{%
(it\omega )^{n}}{n!}\frac{3^{n}-(-1)^{n}}{3\times 4}=\hbar m\omega \beta
\frac{1}{12}(e^{-i\omega t}-e^{3i\omega t}).
\end{eqnarray}%
To be more convincing, we have calculated the next six terms to verify the
correctness of the collection of coefficients. It turns out that those
twelve terms all conform with the series expansion very well. Thus, we get
the time evolution of annihilation operator in Heisenberg picture, that is
Eq. (7).

\setcounter{equation}{0} \renewcommand{\theequation}{B\arabic{equation}}

\section*{Appendix B: Expectation value of quadrature operators}

According to Eq. (7), we have the following expectation values in a coherent
state $\left\vert \alpha \right\rangle $%
\begin{eqnarray}
\left\langle X_{1}^{2}\right\rangle  &=&\frac{1}{4}(1+e^{-2it\omega }\alpha
^{2}+2\alpha \alpha ^{\ast }+e^{2it\omega }\alpha ^{\ast 2})+\frac{1}{48}%
e^{-4it\omega }hm\omega \beta \lbrack -6\alpha ^{4}-6e^{8it\omega }\alpha
^{\ast 4}+  \notag  \label{B1} \\
&&e^{6it\omega }(-6+27\alpha ^{2}+2\alpha ^{3}\alpha ^{\ast }+9\left(
1+4it\omega \right) \alpha ^{\ast 2}+6\left( 1+4it\omega \right) \alpha
\alpha ^{\ast 3}+4\alpha ^{\ast 4}-48\alpha \text{Re}[\alpha ]  \notag \\
&&+e^{2it\omega }(-6+\alpha ^{2}\left( 33-36it\omega -2\alpha ^{2}\right)
-24it\omega \alpha ^{3}\alpha ^{\ast }+3\alpha ^{\ast 2}+2\alpha \alpha
^{\ast 3}+12\alpha \left( -4+\alpha ^{2}\right) \text{Re}\left[ \alpha %
\right]   \notag \\
&&+2e^{4it\omega }(6+5\alpha ^{2}\left( -6+\alpha ^{2}\right) +\alpha ^{\ast
2}(-6-4\alpha \alpha ^{\ast }+\alpha ^{\ast 2}-8\alpha \left( -6+\alpha
^{2}\right) \text{Re}\left[ \alpha \right]   \notag \\
&&+24Abs\left[ \alpha \right] ^{4}\sin ^{2}\omega t)], \\
(\left\langle X_{1}\right\rangle )^{2} &=&\frac{1}{4}e^{-2i\omega t}(\alpha
+e^{2i\omega t}\alpha ^{\ast })^{2}+\frac{1}{24}e^{-4i\omega t}\hbar \omega
m\beta (\alpha +e^{2i\omega t}\alpha ^{\ast })[-3\alpha ^{3}-3e^{6i\omega
t}\alpha ^{\ast 3}  \notag \\
&&+e^{2i\omega t}(6(1-2i\omega t)\alpha \left\vert \alpha \right\vert
^{2}-6\alpha \alpha ^{\ast 2}+\alpha ^{\ast 3}+2(\alpha (6-6i\omega t+\alpha
^{2})-6\text{Re}[\alpha ]))  \notag \\
&&+e^{4i\omega t}(-12\alpha +\alpha ^{3}-6\alpha \left\vert \alpha
\right\vert ^{2}+2\alpha ^{\ast }(6i\omega t+\alpha ^{\ast }(3\alpha
+6i\omega t\alpha +\alpha ^{\ast }))+12\text{Re}[\alpha ])], \\
\left\langle X_{2}^{2}\right\rangle  &=&\frac{1}{4}(1-e^{-2it\omega }\alpha
^{2}+2\alpha \alpha ^{\ast }-e^{2it\omega }\alpha ^{\ast 2})+\frac{1}{48}%
e^{-4it\omega }\hbar m\omega \beta \lbrack (2\alpha ^{4}-3e^{6it\omega
}(\alpha ^{2}-2)  \notag \\
&&+e^{2i\omega t}(6+3\alpha ^{2}(5+12i\omega t)-4\alpha ^{4})+2e^{4it\omega
}(-6-6\alpha ^{2}+\alpha ^{4})+12e^{2it\omega }(e^{2it\omega
}-1)^{2}\left\vert \alpha \right\vert ^{4}  \notag \\
&&+3e^{2it\omega }\alpha ^{\ast 2}(-1-4e^{2it\omega }+e^{4it\omega
}(5-12i\omega t))+2e^{4it\omega }\alpha ^{\ast 4}(e^{2it\omega }-1)^{2}
\notag \\
&&-2e^{2it\omega }\left\vert \alpha \right\vert ^{2}(-12-5\alpha
^{2}-12i\omega t\alpha ^{2}+e^{4it\omega }(\alpha ^{2}-12)+4e^{2it\omega
}(6+\alpha ^{2})+(1+4e^{2it\omega }  \notag \\
&&+e^{4it\omega }(12i\omega t-5))\alpha ^{\ast 2})], \\
(\left\langle X_{2}\right\rangle )^{2} &=&-\frac{1}{4}e^{-2i\omega t}(\alpha
+e^{2i\omega t}\alpha ^{\ast })^{2}+\frac{1}{24}e^{-4i\omega t}\hbar \omega
m\beta (\alpha -e^{2i\omega t}\alpha ^{\ast })[\alpha ^{3}+2e^{2it\omega
}\alpha (3+6i\omega t-\alpha ^{2})  \notag \\
&&+e^{4it\omega }(\alpha ^{2}-6)+6e^{2it\omega }(1+e^{2it\omega }\alpha
^{\ast }(2i\omega t-1))-e^{2it\omega }\alpha ^{\ast 3}(e^{2it\omega }-1)^{2}
\notag \\
&&-6e^{2it\omega }\left\vert \alpha \right\vert ^{2}((-1+e^{2it\omega
}-2i\omega t)\alpha +(-1+e^{2it\omega }(1-2i\omega t))\alpha ^{\ast })], \\
\left\langle \hat{p}^{2}\right\rangle  &=&2\hbar \omega m\left\langle
X_{2}^{2}\right\rangle +\frac{8}{3}\hbar ^{2}\omega ^{2}m^{2}\beta
\left\langle X_{2}^{4}\right\rangle   \notag \\
&=&\frac{1}{2}\hbar \omega m(1-e^{-2i\omega t}\alpha ^{2}+2\left\vert \alpha
\right\vert ^{2}-e^{2i\omega t}\alpha ^{\ast 2})+  \notag \\
&&\frac{1}{24}e^{-4i\omega t}\hbar ^{2}\omega ^{2}m^{2}\beta \lbrack
(6e^{2i\omega t}+6e^{6i\omega t}-3e^{2i\omega t}(3+4e^{2i\omega
t}-12it\omega )\alpha ^{2}+2(3-2e^{2i\omega t}+e^{4i\omega t})\alpha ^{4}
\notag \\
&&+12e^{2i\omega t}(1+e^{4i\omega t})\left\vert \alpha \right\vert
^{4}+16e^{2i\omega t}\left\vert \alpha \right\vert ^{2}(3e^{2i\omega
t}-\alpha ^{2})+(-3e^{2i\omega t}-12e^{4i\omega t}-e^{6i\omega
t}(9+36it\omega ))\alpha ^{\ast 2}  \notag \\
&&-16e^{6i\omega t}\left\vert \alpha \right\vert ^{2}\alpha ^{\ast
2}+2e^{4i\omega t}(1-2e^{2i\omega t}+3e^{4i\omega t})\alpha ^{\ast
4}-2e^{2i\omega t}\left\vert \alpha \right\vert ^{2}(-12(e^{2i\omega
t}-1)^{2}+(-5+4e^{2i\omega t}  \notag \\
&&+e^{4i\omega t}-12it\omega )\alpha ^{2}+(1+4e^{2i\omega t}+e^{4i\omega
t}(-5+12it\omega )\alpha ^{\ast 2})].
\end{eqnarray}

\end{widetext}

\end{document}